\documentclass[a4]{PoS}

\usepackage{amsmath} 
\usepackage{array}
\usepackage{bbold}

\newcolumntype{R}{>{$}r<{$}}
\newcolumntype{C}{>{$}c<{$}}

\title{Symmetry-unrestricted Skyrme mean-field study of heavy nuclei}

\ShortTitle{Symmetry-unrestricted Skyrme-mean-field study of heavy nuclei}

\author{\speaker{W. Ryssens}, P.-H. Heenen\\
        PNTPM, CP229, Universit\'e Libre de Bruxelles, B-1050 Bruxelles, Belgium\\
        E-mail: \email{wryssens@ulb.ac.be}, \email{phheenen@ulb.ac.be}}
\author{M. Bender\\        
        IPNL, Universit\'e de Lyon, Universit\'e Lyon 1, CNRS/IN2P3, F-69622 Villeurbanne, France\\
        E-mail: \email{bender@ipnl.in2p3.fr}\\}

\abstract{ In the light of recent experimental developments, increasing attention is devoted to nuclear phenomena related to rotational excitations of exotic intrinsic nuclear configurations that often lack symmetries present in the majority of nuclei. Examples include configurations with a non-vanishing octupole moment. In order to describe this kind of states, we have developed a new computer code to solve the self-consistent mean-field equations, able to use most of today's effective Skyrme interactions and working in coordinate-space. We report on the development of \texttt{MOCCa}, a code based on the same principles as \texttt{EV8}, but offering the user individual control on many symmetry assumptions. In addition, the HF+BCS pairing treatment of \texttt{EV8} has been generalised to the full machinery of Hartree-Fock-Bogoliubov transformations. We discuss as example the static fission barrier of $^{226}$Ra, prefacing extended studies in the region, using the recent series of Skyrme parameterizations SLy5s1 through SLy5s8. }

\FullConference{54th International Winter Meeting on Nuclear Physics\\
		25-29 January 2016 \\
		Bormio, Italy}

\begin{document}

\section{Introduction: energy density functional theory}
Nuclear structure models employing an energy density functional (EDF) provide a microscopic description of all nuclei over the full nuclear chart~\cite{Bender03}. Several families of EDFs are employed regularly, among which the Skyrme family is the one of interest for this work. Various parameterizations with different aims have been constructed over the past decades, and they have been employed successfully in a multitude of applications, primarily on the mean-field level but also on the beyond-mean-field level.

The self-consistent mean-field or single-reference EDF approach is just one of several possible ways to employ EDFs to study nuclear structure. It enables the user to calculate the properties of the ground state of nuclei along the nuclear chart, and it also allows for studying alternative configurations such as shape isomers or follow the behavior of a nucleus along a rotational band. In this context, several different approaches are (or have been) in frequent use. At the level of the functional, non-relativistic and relativistic EDFs of either zero-range or finite range form are regularly used~\cite{Bender03}. At the level of the variational space, the early Hartree-Fock (HF) calculations are nowadays usually superseded by more general Hartree-Fock-Bogoliubov (HFB) calculations (and their simplification, HF+BCS), but for some applications the HF ansatz remains sufficient~\cite{Shi12,Inakura03}. Finally, several methods to solve the mean-field equations are used, where the main difference is the numerical representation of the single-particle wave functions. The single-particle wave functions are most often represented by an expansion on a truncated basis of oscillator wave functions~\cite{HFODD,Stoitsov13,Carlsson10}, or in a finite volume characterized by some discretization of coordinate space~\cite{Bonche05,Mar14a,Umar1991,PSF08a,Pei2014}\footnote{The differences between both representations are far from trivial, as is evidenced for example by frequently used schemes for the treatment of the pairing interaction and its cutoff, which are often set-up in a manner that is simple in one representation but almost impossible to realize in the other.}.
 
The role of many-body symmetries in the context of solving the mean-field equations is essential. While symmetries convey quantum numbers to different configurations of a nucleus, the variational ansatz (either HF or HFB) introduces the phenomenon of spontaneous symmetry breaking~\cite{RingSchuck}. As an example, consider a calculation that allows for the breaking of spherical symmetry. This calculation explores the degrees of freedom related to deformation of the nuclear density, and consequently is likely to produce a deformed many-body wave function that can be associated with a rotational band. A broken symmetry thus enriches the model, as it allows for the description of a larger range of phenomena.

While breaking many-body symmetries allows the single-reference mean-field method to improve the description of many nuclei, it always comes at both a physical and numerical cost. The numerical cost is related to the enlargement of the variational space: calculations conserving spherical symmetry need only seconds of CPU time, while calculations in 3D for heavy nuclei can take several hours~\cite{Ryssens15}. The physical cost is mostly related to the loss of quantum numbers: a deformed configuration does not have a definite angular momentum quantum number. This severely complicates the interpretation of results, as the experimental levels do carry an angular momentum quantum number. Multi-reference EDF calculations can remedy this problem through the restoration of the broken symmetries. See \cite{Bender03} for an overview and \cite{Bally14} for a recent state-of-the-art example.

The aim of this communication is to present the current state of the development of a new computer code called \texttt{MOCCa}. Its goal is to provide a framework to perform single-reference EDF calculations with a large variety of Skyrme functionals and all of the pairing options (HF, HF+BCS and HFB) while allowing the user full control of the symmetries conserved or broken in the calculation. It is based on the same principles as \texttt{EV8}~\cite{Bonche05,Ryssens15}, representing the single-particle wave functions on a cubic Lagrange mesh~\cite{Baye86,Baye15} in coordinate space. The restrictions on the conserved symmetries are fully controllable by the user and pairing can be treated using the HFB approach. 

In section \ref{section:symmetries} we detail the symmetry groups of interest, as well as the currently available options in \texttt{MOCCa}. As an example, section \ref{section:Radium} will detail the description of the fission barrier of $^{226}$Ra using the recent SLy5sX family of functionals, illustrating the power of symmetry breaking.

\section{Symmetry conservation and breaking: \texttt{MOCCa}}
\label{section:symmetries}
Employing a cubic Lagrange mesh~\cite{Baye15} as done in \texttt{MOCCa} and \texttt{EV8} provides a natural way for any calculation to break rotational symmetry. Most deformed configurations of interest, however, keep some remaining symmetries, for example related to reflections of some kind or to rotations about a specific axis and angle.
This can be exploited to reduce the complexity and cost of the calculations of such states, as each remaining symmetry brings some simplification to the equations of motion to be solved. Indeed, as long as the numerical representations used are strictly equivalent, calculating a state that adopts certain symmetries with a symmetry-unrestricted code yields exactly the same result as a calculation with a code that imposes the same symmetries on the calculation from the beginning. For the purpose of describing deformed nuclei, the groups that cover the majority of cases of interest are obtained by combining the spatial rotations, plane and point reflections that map a rectangular box onto itself with time reversal. Two distinct groups have to be distinguished, $D^{T}_{2h}$ for systems with an even number of particles, and $D^{TD}_{2h}$ for systems with an odd particle number \cite{Doba00a,Doba00b}. They are composed of the following many-body operators
\begin{eqnarray}
  D^{T}_{2h} &=& \left\{ \hat{\mathbb{1}}, \hat{P}, \hat{T},\hat{P}^T, \hat{R}_{\mu}, \hat{R}^T_{\mu}, \hat{S}_{\mu}, \hat{S}^T_{\mu}  \right\} \, , \\
  D^{TD}_{2h} &=& \left\{ \hat{\mathbb{1}}, -\hat{\mathbb{1}}, \hat{P},-\hat{P}, \hat{T},-\hat{T}, \hat{R}_{\mu},-\hat{R}_{\mu}, \hat{S}_{\mu}, -\hat{S}_{\mu}, \hat{S}^T_{\mu}, -\hat{S}^T_{\mu} \right\}  \, , 
\end{eqnarray}
where the Greek index $\mu$ ranges over the three Cartesian directions $x,$ $y$ and $z$. The operator $\hat{P}$ is the parity operator, while $\hat{T}$ is the time-reversal operator\footnote{Note that the operator $-\hat{\mathbb{1}}$ needs to be introduced explicitly into the group strcuture of $D^{TD}_{2h}$, since $\hat{T}^2 = - \hat{\mathbb{1}}$ for odd systems. This accounts for the appearance of operators with a minus sign such as the pair $(\hat{P}, -\hat{P})$.}. The three operators $\hat{R}_{\mu}$ are the signature operators defined as 
\begin{equation}
  \label{Symmetries:eq:sigalt}
\hat{R}_{\mu} = \exp \left[ -i \frac{\pi}{\hbar} \hat{J}_{\mu} \right] \, , 
\end{equation}
where $\hat{J}_{\mu}$ is the many-body angular momentum operator. The signature operators thus represent rotations about the Cartesian axes of 180 degrees. The group structure is completed by the following identities
\begin{eqnarray}
  \hat{P}^T       &=& \hat{P} \hat{T}  \, , \\
  \hat{S}_{\mu}   &=& \hat{P} \hat{R}_{\mu}  \, , \\
  \hat{S}^T_{\mu} &=& \hat{T} \hat{S}_{\mu} \, .
\end{eqnarray}
All of these many-body operators are product of the corresponding single-particle operators, meaning that conserved many-body symmetries do not carry-over to single-particle symmetries on a one-to-one basis. This is also at the heart of the difference in group structure between $D^{T}_{2h}$ and $D^{TD}_{2h}$: it is due to the  bosonic nature of even systems, while odd systems behave like fermions. The behavior of time-reversal is the best example\footnote{Note that the multiplication rules for the signature operators $\hat{R}_{\mu}$ also differ between even and odd systems, see~\cite{Doba00a, Doba00b}. This is however of no consequence for this communication.}: $\hat{T}^2 = \pm \mathbb{1}$ depending on whether the number of nucleons is even (+) or odd (-). This means that, while both groups are similar superficially, $D^{T}_{2h}$ and $D^{TD}_{2h}$ exhibit different multiplication tables.

The group $D^{T}_{2h}$ for even-even nuclei is most easily interpreted as the classical group of three plane symmetries, enriched by the time-reversal operator $\hat{T}$. For a given mean-field calculation of a nucleus, any subgroup of $D^{T}_{2h}$ could be designated as a set of conserved symmetries. The discarded symmetry operators then make up the set of potentially broken symmetries: the mean-field configuration does not need to respect them, thereby enlarging the variational space. The code \texttt{EV8} for example, conserves the entire group $D^{T}_{2h}$ and thus allows for a single combination of broken symmetries taken from $D^{T}_{2h}$: the empty set $\emptyset$. Two other unpublished codes that use the Lagrange mesh representation allow for a different choices: breaking parity~\cite{Bender04} or time-reversal symmetry~\cite{Bonche87}.

By contrast, the \texttt{MOCCa} code allows the user to conserve many subgroups of $D^{T}_{2h}$ when describing nuclei with an even number of particles. While not all possible subgroups are currently implemented, the current version does provide the choice of 16 subgroups that can be conserved, thus exhausting a large part of all possibilities. These possibilities include both the empty group and the full $D^{T}_{2h}$ group, the former corresponding to a completely symmetry unrestricted calculation and the latter to the case of \texttt{EV8}. In addition, the full HFB ansatz is available for all of these subgroups, where \texttt{EV8} was limited to HF and HF+BCS calculations. Combined with the symmetry options, this gives \texttt{MOCCa} a significantly larger range of applicability. 
\section{The fission barrier of $^{226}$Ra}
\label{section:Radium}
\begin{figure}
\center
\includegraphics[width=0.99\textwidth]{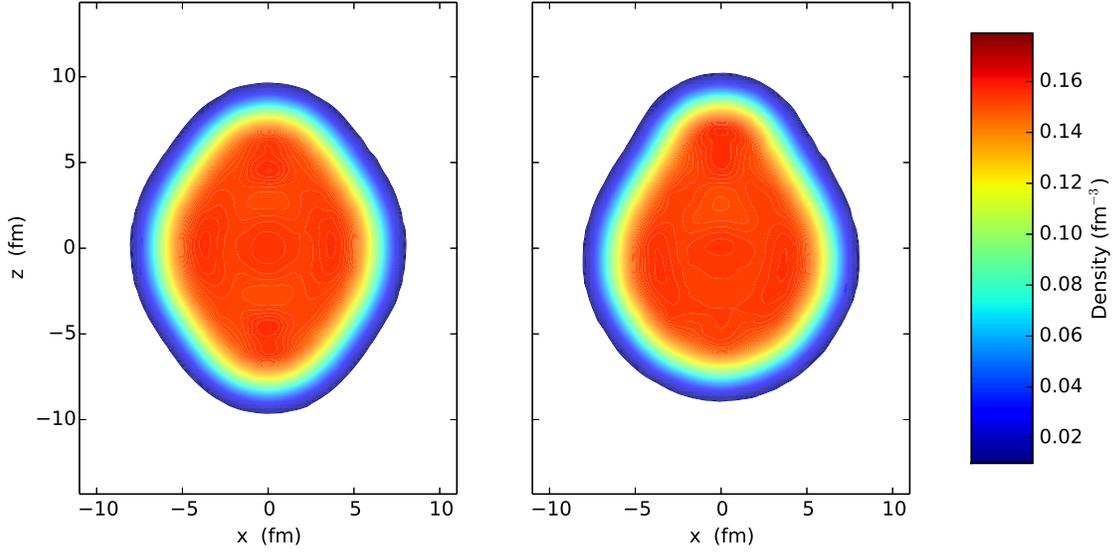}
\caption{Density contours in the x-z plane for two configurations of $^{226}$Ra using the SLy5s1 parameterization. On the left, a prolate configuration that conserves parity, on the right an octupole deformed state. }
\label{fig:densities}
\end{figure}
As an example of the possibilities of the framework, we describe the static fission barrier of $^{226}$Ra, using the recent family of fits SLy5s1 to SLy5s8~\cite{Jodon16}. These parameterizations have been adjusted with the same procedure as SLy5*~\cite{Jodon16}, but with an extra constraint on the surface properties: going from SLy5s1 to SLy5s8, their surface tension increases in equal steps. As we will see, the breaking of both axial symmetry and parity will be important for a proper description of this nucleus, giving rise to triaxial and octupole deformed configurations, respectively. As examples for the relevant configurations, consider Fig.~\ref{fig:densities}. Density contours in the x-z plane are shown for a prolate and an octupole deformed configuration, respectively. 
\begin{figure}[t!]
\center
\includegraphics[width=.95\textwidth]{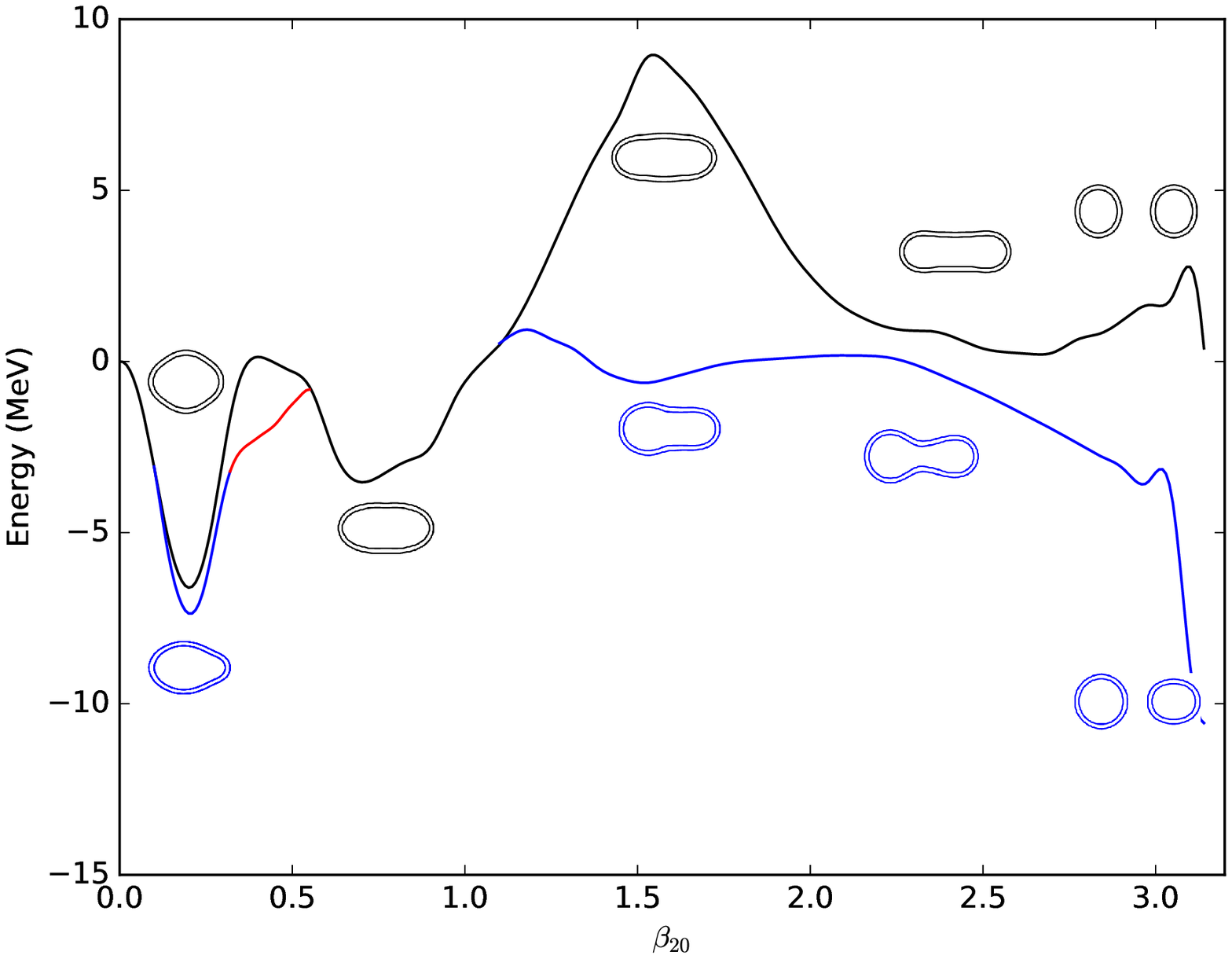}
\caption{Deformation energy of $^{226}$Ra as a function of the deformation parameter $\beta_{20}$, calculated with the SLy5s1 parameterization. Black lines indicate axial, parity conserving configurations while red lines indicate triaxial, parity conserving configurations. Finally, blue lines indicate axial configurations with a non-zero octupole moment. Density contour plots sketch the type of configurations encountered along the energy curve. All energies are relative to the spherical configuration.}
\label{fig:barrier}
\end{figure}
 Pairing was treated using the full HFB ansatz, using a density-dependent zero-range pairing interaction as described in~\cite{Krieger90}. A pairing cutoff was introduced, symmetric around the Fermi energy at 5 MeV~\cite{Ryssens15}. In addition, the Lipkin-Nogami (LN) prescription~\cite{Gall94} was employed to ensure the presence of pairing for all nuclear configurations. The pairing strengths $(V_n, V_p)$ used are 
\begin{equation}
  (V_n, V_p) = (1000,1450) \text{ MeV fm}^3 \, ,
\end{equation}
which have been determined by adjusting to the three-point proton and neutron gap of $^{216}$Ra for the SLy5s1 parameterization. The calculations for the even nuclei used in the adjustment conserved time-reversal symmetry as well as parity, but allowed for triaxial deformations. For the even-odd nuclei in the adjustment time-reversal invariance was broken and HFB quasiparticles corresponding to the experimental ground states were self-consistently blocked~\cite{Heenen95}. 

The fission path, as a function of the deformation parameter $\beta_{20}$, is shown in Fig.~\ref{fig:barrier} for the SLy5s1 parameterization with HFB+LN pairing as described before. The different colors represent calculations with a different set of imposed symmetries: the black curve is calculated while imposing both axial and reflection (parity) symmetry. The red curve corresponds to lower solutions that are found around the first barrier when relaxing the constraint on axial symmetry. All of these conserve parity. The blue curve indicates lower solutions that are found when relaxing parity. Density contours in the x-z plane are included along the barrier to give the reader a visual representation of the type of shapes encountered.

The absolute minimum found at $\beta_{20} \approx 0.2$ is octupole deformed, a particularity of nuclei in this mass region, while the parity-conserving minimum of the black curve is actually a saddle point. These minima are the configurations for which Fig.~\ref{fig:densities} shows the density distribution in the x-z plane. However, 
the energy surface is rather flat in this region and octupole deformation lowers the energy by slightly more than one MeV. At the first barrier  at $\beta_{20} \approx 0.4$ $^{226}$Ra, gradually loses its octupole deformation. Nevertheless, as mentioned before, triaxial degrees of freedom significantly lower the barrier height. The configuration corresponding to the broad second minimum at $\beta_{20} \approx 0.7$ takes an axial and parity conserving shape. If octupole deformations are not allowed at larger deformations, one finds a broad and very high second barrier around $\beta_{20} \approx 1.5$ that is followed by a third barrier at about $\beta_{20} \approx 3.0$ before the calculation jumps to a solution with two separate identical fragments. The two outer barriers are substantially lowered when allowing reflection asymmetric shapes as indicated by the blue path in Fig.~\ref{fig:barrier}. The resulting height of the asymmetric barrier of 8.31 MeV agrees well with the experimental value of 8.5 MeV as reported in the RIPL-3 database \cite{RIPL}. This fission path leads to an asymmetric split-up, with the larger fragment being near-spherical and the smaller one remaining deformed. We have not found any solution that is non-axial with finite octupole moments. All calculations initialized with such shapes converged to states for which one or the other of these symmetries is reestablished.

\begin{figure}[t!]
\center
\includegraphics[width=.99\textwidth]{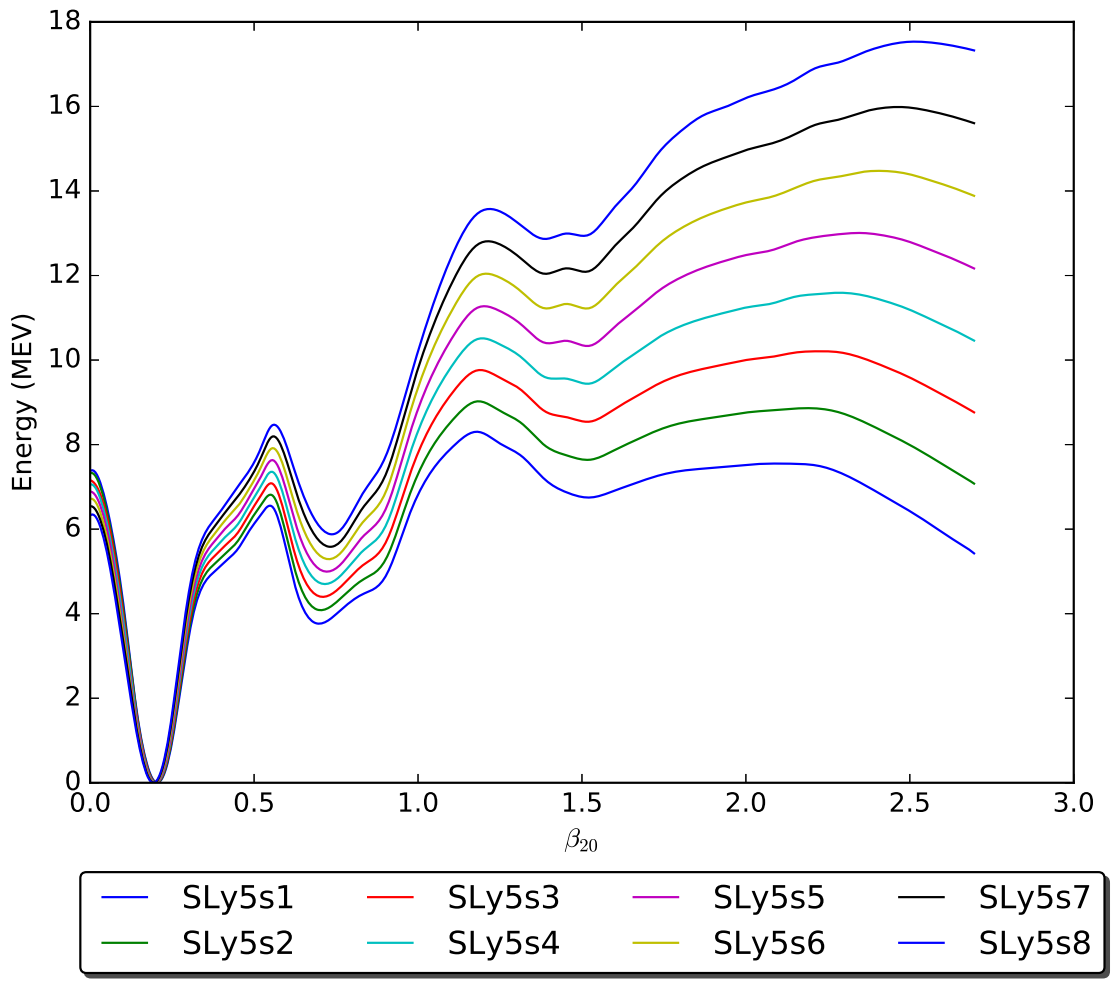}
\caption{Fission barrier of $^{226}$Ra as a function of $\beta_{20}$ for the SLy5sX functionals with $X=1,\ldots,8$ with HFB+LN pairing.  Energy is relative to minimum of the deformation surface.}
\label{fig:barriers}
\end{figure}
It is interesting to compare to previous results of~\cite{Rutz95}, obtained within the framework of relativistic mean-field theory and incorporating HF+BCS pairing. Note that the older calculation does not include triaxial degrees of freedom. Both calculations are in rather good qualitative agreement obtaining minima at comparable quadrupole and octupole deformations. One notable difference is the alternative branch along the symmetric fission path reported in~\cite{Rutz95}, differing mainly in hexadecapole deformations, that we have not found. A second difference concerns the relative energy of the symmetric and asymmetric fission paths at large deformations ($\beta_{20} \approx 2.5$): while the asymmetric path is preferred by the calculation shown here, the symmetric path is favoured in~\cite{Rutz95} in a small deformation region before the onset of the third barrier. These differences between the energy landscapes can be attributed to slight differences in the properties of the effective interactions used in both cases, in particular concerning shell structure and its change with deformation. There is however no fundamental difference between both approaches.


Note that it cannot be assumed that the symmetry-restricted energy curve (the black line in Fig.~\ref{fig:barrier}) corresponds to a physical fission path with higher symmetries, i.e.\ that it is a local minimum in all non-constrained multipole  degrees of freedom. In general, only the much more time-consuming calculation of a multi-dimensional energy surface can answer this question. Similarly, multi-dimensional calculations are sometimes needed to find the correct height of saddle points when the calculated fission path jumps from one valley to another, which is signaled by discontinuities in the non-constrained multipole moments. The calculation of deformation energy curves as presented here, however, is sufficient to analyze the overall structure of a fission barrier and to identify the relevant shape degrees of freedom for such more detailed studies. Such multi-dimensional studies are planned for the near future, and \texttt{MOCCa} is capable to carry them out.

For the excellent reproduction of the barrier height of $^{226}$Ra reported above it is not sufficient to have numerical codes that cover all relevant degrees of freedom. First and foremost it requires a properly fine-tuned parameterization of the energy density functional. As already mentioned, the SLy5s1 parameterization is one out of a series of eight fits that differ in the value of their surface energy coefficient $a_{\text{surf}}$ \cite{Jodon16}. This quantity provides a measure of the surface tension of symmetric nuclear matter. While the complex topography of an energy surface such as the one displayed in Fig.~\ref{fig:barrier} is determined by changes of shell structure with deformation, the surface tension provides the smooth backdrop on which  these shell effects generate valleys and ridges. In a simple liquid-drop picture of the nucleus, deforming a spherical nucleus leads to an energy loss that equals $a_{\text{surf}}$ times the change of the nucleus' surface. The balance of this effect with the gain in energy from reducing the Coulomb repulsion by deforming the nucleus then leads to a broad singly-humped macroscopic barrier. The systematic variation of $a_{\text{surf}}$ in steps of 0.2 MeV provided by the SLy5sX family of fits can be used to illustrate the interplay of these microscopic and macroscopic effects.

Figure~\ref{fig:barriers} shows the deformation energy curve of $^{226}$Ra for all eight parameterizations from the SLy5sX series. Pairing correlations are treated as before for all of them. Strikingly, all curves have the same overall shape, which indicates that the shell effects do not significantly change when going from one parameterization to another. This is also corroborated by the fission path being the same for all parameterizations: all multipole deformations of degree $\ell = 2$, 3, and 4 are virtually identical along the entire energy curve. What does dramatically change is the overall slope of the energy curves. Indeed, with increasing deformation the contribution of the surface energy grows quickly, such that, when going from the energy curve obtained for SLy5s1 to the one for SLy5s8, each is systematically higher than the previous one. Comparing the two extremes, SLy5s1 and SLy5s8, the height of the first barrier is already affected at a level of 1.5 MeV, while the excitation energy of the second minimum increases by about 3 MeV. Even more dramatically, the height of the second barrier grows by about 6 MeV, and the third barrier by even more than 9 MeV. Most strikingly, the third barrier, relatively unpronounced for SLy5s1 and SLy5s2, becomes the dominant one for the other fits as its height increases much more quickly with $a_{\text{surf}}$ than the one of the second barrier, see
Table~\ref{tab:barriers}.

The same qualitative and quantitative variation of barrier heights and excitation energies of shape isomers is also found in similar calculations for other nuclei \cite{Jodon16}, indicating that the surface tension of SLy5s1 is more realistic than the one of the other fits from the same series when, as done here, one aims at mean-field calculations of energy surfaces without any quantum corrections related to deformation degrees of freedom. The most relevant one of these for the present discussion is projection of angular momentum, either exactly in a multi-reference framework~\cite{Bender04}, or in an approximate way through a rotational correction. Its effect tends to increase with deformation \cite{Bender03}. In order to reproduce data, calculations with such correction require parameterizations with larger surface tension than calculations without it. Thus, the possibility of achieving a parameter fit of an EDF that describes fission barriers at any level of modeling is very unlikely. Still, with the protocol suggested in \cite{Jodon16} it is possible to construct parameterizations with controlled surface properties that can be adapted to the context of their use without any difficulty.

\begin{table}[t!]
  \center
\begin{tabular}{lccc}
  \hline
  \hline
  Functional  & $a_{\text{surf}}^{\text{MTF}}$ (MeV) & Barrier height (MeV) & Barrier location $\beta_{20}$ \\
  \hline
SLy5s1  &  18.0 &8.31   &  1.19 \\
SLy5s2  &  18.2 &9.03   &  1.19 \\
SLy5s3  &  18.4 &10.21  &  2.23 \\
SLy5s4  &  18.6 &11.59  &  2.29 \\
SLy5s5  &  18.8 &13.01  &  2.35 \\
SLy5s6  &  19.0 &14.47  &  2.41 \\
SLy5s7  &  19.2 &15.99  &  2.47 \\
SLy5s8  &  19.4 &17.53  &  2.52 \\
  \hline
\end{tabular}
\caption{Surface energy coefficients $a_{\text{surf}}^{\text{MTF}}$, highest energy and its location along the fission barrier of $^{226}$ Ra for SLy5sX series of fits. The surface energy coefficients given are calculated in Modified Thomas-Fermi approximation as used during the parameter adjustment, see~\cite{Jodon16} for values obtained in other frameworks.}
\label{tab:barriers}
\end{table}

\section{Conclusion}

We have reported on the development of the computer code \texttt{MOCCa} for self-consistent mean-field calculations with non-relativistic energy density functionals, which is a direct generalization of the methods employed in \texttt{EV8}. Benefiting from the accuracy of a Lagrange mesh representation~\cite{Ryssens15b}, the code is currently capable of performing calculations respecting the symmetries of a large number of subgroups of $D^{T}_{2h}$ (for even-A nuclei) and $D^{TD}_{2h}$ (for odd-A nuclei). To the best of our knowledge, the only other code that offers a comparable degree of flexibility at this point in time is HFODD~\cite{HFODD}, which differs from the approach presented here by representing the single-particle wave functions on a truncated harmonic oscillator basis.

We have shown here the application of the framework to the fission barrier of $^{226}$Ra using the recent set of SLy5sX parameterizations, pioneering with this example the description of triaxial and octupole deformed states using HFB pairing on a Lagrange mesh. Further studies for other nuclei, for which additional shape degrees of freedom that were not found to be relevant for $^{226}$Ra might play a crucial role, are ongoing.

\section*{Acknowledgments}

This work has been supported by the Belgian Office for Scientific Policy under Grant No. PAI-P7-12. Part of the computations was performed using the Plateforme Technologique de Calcul Intensif (PTCI) located at the University of Namur, Belgium, which is supported by the F.R.S.-FNRS under the convention No. 2.4520.11. Another part of the computations was performed on the HPC clusters HYDRA and VEGA hosted at the Computing Centre, an IT service co-funded by the Vrije Universiteit Brussel and the Universit\'e Libre de Bruxelles. Both the PTCI and the Computing Centre are members of the Consortium des \'Equipements de Calcul Intensif (C\'ECI).

\end{document}